\documentclass[9pt,twocolumn,twoside]{osajnl}

\journal{josab} 

\setboolean{shortarticle}{false}

\title{Dual-Core Optical  Fibers for  Efficient Mid-Infrared Generation via  Third Order  Frequency Mixing and Coupling-Length Phase Matching}
\author[1]{Jing Su}
\author[1,*]{Ivan Biaggio}
\affil[1]{Department of Physics and Center for Photonics and Nanoelectronics, Lehigh University, Bethlehem, PA 18015}
\affil[*]{Corresponding author: biaggio@lehigh.edu}

\dates{Compiled \today}

\ociscodes{(160.4330)    Nonlinear optical materials;  (190.4360)   Nonlinear optics, devices}

\doi{\url{http:\\dx.doi.org/xxxx/xx.XX.XXXXXX}}

\begin{abstract}
Appropriately designed dual-core fibers  using coupling-length phase matching (CLPM)  allow  for phase-matched  frequency downconversion over wide frequency intervals using the  third-order optical nonlinearity of glass.   By tuning the distance between the two cores, CLPM allows  continuously tunable phase matching for widely different wavelengths for the process in which a pump wave at a frequency $\omega_2$ generates or amplifies two waves with frequencies $\omega_3>\omega_2$ and $\omega_1 = 2 \omega_2 - \omega_3$. The intensity-dependent correction that accounts for nonlinear phase-modulation  is derived in general. In addition,  a specific CLPM configuration  is found to be insensitive to phase-modulation, and can  achieve 100\% theoretical quantum yield for pump wave injection in  one core. Fiber-based frequency converters can thus be designed for large differences between pump wavelengths and generated wavelengths, with the phase-matched interaction enabling the use of  meter-long fibers to compensate for  low third-order susceptibilities. Examples of fiber designs for pump wavelengths at 1.3 and 1.55 $\mu$m, to generate radiation with wavelengths longer than 2 $\mu$m, are discussed for silica and fluoride fibers.
\end{abstract}

\setboolean{displaycopyright}{true}

\begin{document}

\maketitle

\section{Introduction}
The coupling between the cores in dual core optical  fibers can compensate for the natural bulk dispersion of the material and lead to flexible phase matching opportunities for third-order nonlinear optical interactions. The availability of appropriate phase-matching schemes  enables the design of fiber-based frequency converters where a relatively weak third-order susceptibility is compensated by a long interaction length, an attractive opportunity for the implementation of new fiber-based optical frequency converters. 

In this work, we focus on  the third-order nonlinear optical interaction of three waves with frequencies $\omega_3>\omega_2>\omega_1$ that satisfy $ \omega_1 +  \omega_3= 2 \omega_2$. Such an interaction can be used to create shorter wavelengths from longer wavelengths, or vice-versa. We will almost exclusively use the example of frequency downconversion,  the generation of  longer wavelengths from visible or near-infrared wavelengths,  but all our results equally apply to any other frequency conversion scheme using this  third-order three-wave mixing process. 

Third-order three-wave mixing can be seen  as  a   difference frequency generation process ($2 \omega_2 - \omega_3 \to \omega_1$), or also a  parametric  process ($2 \omega_2  \to \omega_1 + \omega_3$).  These effects are the third-order equivalents of the second-order nonlinear optical interactions used in  difference frequency generation, optical parametric generation (OPG), optical parametric oscillation (OPO), parametric fluorescence, or parametric down-conversion. In the following we will generally describe the wave at frequency $\omega_2$ as the ``pump'' wave, and the waves at $\omega_3$ and $\omega_1$ as the ``generated waves''. Either  of the two generated waves can also play the role of a ``seed wave'' when it is injected into the system together with the pump wave. Whenever we will discuss this possibility, we will do so by considering the wave at $\omega_3$ as the seed wave, for generation into the farther infrared. Such fiber-based frequency converters would be useful to bridge gaps in the availability of laser sources in farther infrared, for example in coincidence with atmospheric transmission windows, where lasers could be employed for infrared backscatter imaging spectroscopy \cite{Breshike19}, LIDAR, or other investigations. As a third-order equivalent to parametric fluorescence they could also be developed into a source of photon pairs and for heralded single photon emission \cite{Rottwitt18,Paesani20}.

While three- and four-wave mixing in fibers has been treated extensively in the literature \cite{garth1986,sammut1989,inoue1992,inoue1994,aso1999,Yamamoto97,Song99,Serkland99,Harvey03,Wong07,Herzog12,Agrawal19book}, that previous work mostly  concentrated on the case of small frequency differences, which enabled natural phase matching for the above process, or the simplified analysis of phase-mismatch by series expansion near the degeneracy point where all frequencies are equal. The  case of large frequency differences between the interacting waves, which is the focus of this work, has been treated less often \cite{Lin81,Lin82,Lin83,Agrawal19,wong2007} and depends on the features of the refractive index dispersion over a wide range of wavelengths. In this case, the   coherence length for phase-matched three-wave mixing is in general much shorter  than the length required to obtain a good conversion efficiency. 

Below, we  first present a  short review of the third-order three-wave mixing process in bulk glasses and in single-mode fibers, which we will use to  establish the description of the physical effects that come into play, and all the necessary definitions. As part of this short review we will show  how   glasses with normal dispersion  always enable a particular choice of widely separated wavelengths where the three-wave mixing described above is phase matched. This particular phase-matching condition can also be realized in single-core fibers \cite{Lin81,Lin82,Lin83} but it is constrained to a small interval of pump wavelengths,  with   a limited range of wavelengths for the  generated waves, which can only be weakly influenced by the two available design parameters, the refractive index contrast between core and cladding and the core radius. Other more general phase matching possibilities for this three-wave mixing scheme could in principle include spatial structuring of the material along the propagation length, but this solution would be very inefficient, and difficult to control over long propagation lengths.

Next, we will discuss the use of dual-core, or twin core, fibers  \cite{Peterka00,Zou13,Zhao21} with large coupling between the two cores  to achieve the necessary design flexibility and obtain phase-matched frequency conversion  for  a wide range of pump wavelengths and generated wavelengths. The effect of the distance between the cores can be understood in terms of coupling-length phase matching (CLPM) \cite{Biaggio14}, or in terms of intermodal coupling between the nearby supermodes of the dual core waveguide. The use of two twin cores effectively  multiplies the possibilities to achieve phase matching. This then allows to choose a large variety of pump wavelengths, with increased flexibility in tuning the design for different output wavelengths.

Finally, we must stress that the frequency conversion process that we we are considering is characterized by interacting waves with widely different wavelengths, and by an interaction length required for significant frequency conversion that is of the orders of meters or fractions of a meter.  These relatively large conversion lengths, determined by the strength of the third-order nonlinearity and available intensities,  are enabled by phase matching. They are naturally not an issue in optical fibers at wavelengths where the absorption lengths are  longer than the conversion lengths, and for interacting waves that do not consist of pulses so short that they would walk away from each other over a conversion length because of different group velocities. Because of this, we will analyze this interaction in a continuous wave limit that, given the dispersion of glass and the interaction lengths we consider, will be valid for pulse  durations of nanoseconds or longer. The field of application of the type of frequency converter that we propose in this work is that of small bandwidth lasers with longer pulses.

\begin{table}[b!]
\caption{Sellmeier parameters used in this work for SiO$_2$, ZrF$_4$, and InF$_3$ glasses. The silica parameters are those for the 8655 and 7979 Corning glass. The fluoride parameters are for the  cladding of Thorlabs single-mode fluoride  patch cables. } 
\centering 
\begin{tabular}{c c c c c c c c c } 
\hline\hline 
 & SiO$_2$   &  ZrF$_4$ & InF$_3$   \\ [0.5ex] 
\hline
$a_1$ & 2.623483282E$-$2 & 0.705674 & 0.68462594 \\
$a_2$ & 7.306029048E$-$1 &0.515736  & 0.4952746 \\
$a_3$ & 3.475321572E$-$1 & 2.204519 & 1.4841315 \\
$a_4$ & 9.216052441E$-$1 & 0.0 & 0.0 \\
$b_1$ & $-$5.783959035E$-$3 & 0.087503 &  0.0680833 \\
$b_2$ & 5.600103210E$-$3 & 0.087505 & 0.11054856 \\
$b_3$ & 1.389808930E$-$2 & 23.80739 &  24.4391868 \\
$b_4$ & 1.006578079E$+$2 & 0.0 & 0.0 \\
\hline 
\end{tabular}
\label{sellmeierParams} 
\end{table}

\subsection{Background and nomenclature}
We will  develop examples of phase-matched three-wave interaction in step-index cylindrical waveguides, both as a single waveguide, as well as two parallel waveguides. In all cases, every single waveguide will operate in its lowest-frequency fundamental mode, and we will rely on the standard expressions for the parameters of  cylindrical waveguides to obtain the  propagation constants and coupling coefficients, \cite{Marcuse74,Snyder83}
\begin{eqnarray}
\Delta &=& \frac{n_{core}^2-n_{clad}^2}{2 n_{core}^2} \\
V &=& k_0 n_{core} a \sqrt{2 \Delta} \\
n_{\rm eff} = \frac{\beta}{k_0} &=& \sqrt{n_{clad}^2 + b (n_{core}^2-n_{clad}^2)} \label{nEff} \\
u &=& a \sqrt{k_0^2 n_{core}^2 - \beta^2} \\
w &=& a \sqrt{ \beta^2 - k_0^2 n_{clad}^2 } \\
\kappa &=& \frac{\sqrt{2 \Delta}}{a} \frac{u^2}{V^3} \frac{K_0(w d/a)}{K_1(w)^2} \label{kappaEq}
\end{eqnarray}
where $k_0=\omega/c=2 \pi/\lambda$, with $\lambda$ the vacuum wavelength of the electromagnetic wave. The step-index cylindrical waveguide is described by its core radius, $a$, and by $n_{core}$ and $n_{clad}$, its core and cladding refractive indices. To simplify our treatment, we will generally consider a constant, wavelength-independent, small refractive index contrast $\Delta n = n_{core} - r_{clad}$ that gives $\Delta \ll 1$. The other parameters listed above are the normalized frequency-parameter $V$,  the propagation constant for the lowest guided mode $\beta$, its effective refractive index $n_{\rm eff}$, and  the normalized propagation constant $b=b(V)$ obtained from the $V$-parameter by solving the characteristic transcendental equation for the lowest frequency guided mode  \cite{Marcuse74}.  We will also rely on the approximate expression (\ref{kappaEq}) for the coupling coefficient $\kappa$ between the fundamental modes in two cylindrical waveguides with a center-to-center distance $d$. This expression is valid for larger values of $w d/a$ (See Ref.~\cite{Snyder83}, Chapter 18). Whenever necessary, we have supplemented the results obtained in this way by also calculating the propagation constants of the modes using COMSOL Multiphysics software \cite{COMSOL}.

For the examples presented in this work, we will consider  silica and fluoride glasses with refractive index dispersion described by the standard Sellmeier expression $n^2 - 1 = \sum_{i=1}^4 a_i/(1-b_i^2/\lambda^2)$ with the wavelength $\lambda$ in $\mu$m and the parameters $a_i$ and $b_i$ given in Table \ref{sellmeierParams}.

In the third-order three-wave mixing process, the power of the pump wave at $\omega_2$ is transferred to the generated waves at $\omega_1<\omega_2$ and $\omega_3>\omega_1$. This is described by the third-order susceptibility $\chi^{(3)}_{ijkl}(- (2 \omega_2 - \omega_3),\omega_2, \omega_2, -\omega_3)$. The corresponding nonlinear optical polarization that leads to the generated wavelengths is
\begin{equation}
P_{NL}^{(\omega_1)}  =  \tfrac{3}{4}\epsilon_{0}\chi^{(3)}_{\rm eff}(- \omega_1,\omega_2, \omega_2, -\omega_3)[E^{(\omega_2)}] ^{2}E^{(\omega_3)^{*}} \label{nlopol}
\end{equation} 
where the $ E^{(\omega_{i})}$ are the amplitudes of the interacting optical electric fields, defined as the real part of their complex representation, the superscript asterisk stands for complex conjugation, and $\chi^{(3)}_{\rm eff}$ is the coefficient of the third-order susceptibility tensor that in general depends on the polarization of the interacting waves. In silica glass,  $\chi^{(3)}_{\rm eff} = \chi^{(3)}_{1111} \sim 2 \times 10^{-22} {\rm V}^2 {\rm m}^{-2}$ for interacting waves polarized parallel to each other \cite{Milam98}. We will use  S.I. units throughout.

\section{Description of three-wave interaction}

For the following analysis, we  define the amplitudes
\begin{equation}
A_i =  \sqrt{\dfrac{\epsilon_0 c n_i}{\hbar \omega_i}} E^{(\omega_i)}  \label{amplitudesDef}
\end{equation}
where $E^{(\omega_i)}$ is the amplitude of the optical electric field, $ n_i $ is the effective refractive index for the wave with frequency $ \omega_i $, $\epsilon_0$ is the permittivity of vacuum,  $c$ is the speed of light, and $\hbar$ is the Plank constant. The $A_i$ have units of  m$^{-1}$s$^{-1/2}$and their square represents a photon flux, which will become useful when considering the efficiency of this interaction in terms of the quantum yield introduced below.

We also define an effective nonlinear optical coefficient
\begin{equation}
\chi  =  \dfrac{3}{8} \chi_{\rm eff}^{(3)}  \frac{\hbar}{\epsilon_0 c^2} \sqrt{\dfrac{\omega_{1}\omega_{2}^{2}\omega_{3}}{n_{1}n_{2}^{2}n_{3}}} S , \label{chidef}
\end{equation}
whose units are the product of length and time (m s). Here,  $S$ represents the  overlap integral between the transverse mode-profiles of the interacting waves for propagation in a waveguide, with $S = 1$ for plane-wave interaction. For this work, $S$ does not play a big role:   third-order nonlinear optical interactions   happen equally well  both in the core and in the cladding region, and  $S$ simply decreases slightly from unity as the wavelength difference between the waves increases.  

Finally, we will need to  analyze the efficiency of the frequency down-conversion process. 
Since conversion efficiency defined in terms of power of the output with respect to the input power depends on the frequencies,  conversion efficiency is best discussed in terms of a quantum yield.  The nonlinear optical interaction described above corresponds to the annihilation of two photons in the pump wave at frequency $\omega_2$ and the creation of one photon each in the two waves at frequencies $\omega_1$ and $\omega_3$. Assuming that the initial amplitude of the generated wave is  $A_1(0) = 0$, the quantum yield of this process is
\begin{equation} 
\phi(z) =  \frac{2 |A_1(z)|^2}{|A_2(0)|^2} . \label {quantumyield}
\end{equation}
This is effectively a  ``photon conversion efficiency'' that reaches $\phi=1$ when the pump wave is fully depleted. It is obtained by  dividing the total number of photons that went into the generated wavelengths (proportional to $2|A_1(z)|^2$), by the initial number of photons in the pump wave (proportional to $|A_2(0)|^2$). We will show that for the three-wave mixing process discussed here it is always possible to take into account both phase-matching and the effect of self- and cross-phase modulation to reach a quantum yield of $\sim 100$\% after a sufficiently large interaction length. We note that the quantum yield that we define here must be seen as a theoretical figure of merit for the corresponding process, a fundamental upper limit for the efficiency that can be reached in practical implementations.

\subsection{Single mode wave interaction} 
This section considers three-wave interaction for single-mode propagation, as can be found both for free space propagation, or in a single-mode waveguide, and will then be followed by a discussion of dual-core waveguides. \label{singlemodeinteraction}

Inserting the time- and space-dependence of the three  propagating waves and the induced nonlinear optical polarization (\ref{nlopol}) into the wave equation  leads to the following set of coupled-wave equations for the amplitudes (\ref{amplitudesDef}),
\begin{eqnarray}
\dfrac{\partial}{\partial z}A_{1} & = & i \: \chi [A_{1}(2 A_{3}A^{*}_{3} + 2 A_{2}A^{*}_{2} + A_{1}A^{*}_{1}) \nonumber \\ && + [A_{2}]^{2}A^{*}_{3} e^{i \Delta k z}]   ,      \label{cw1single} \\
\dfrac{\partial}{\partial z}A_{2} & =& i \: \chi [A_{2} (2 A_{3}A^{*}_{3} + A_{2}A^{*}_{2} + 2 A_{1}A^{*}_{1}) \nonumber \\ && + 2 A_{1}A^{*}_{2}A_{3} e^{-i \Delta k z}]   ,     \label{cw2single} \\
\dfrac{\partial}{\partial z}A_{3} & =& i \chi [A_{3}(A_{3}A^{*}_{3} + 2 A_{2}A^{*}_{2} + 2 A_{1}A^{*}_{1}) \nonumber \\ && + [A_{2}]^{2}A^{*}_{1} e^{i \Delta k z}]  ,  \label{cw3single} 
\end{eqnarray}
where   $\Delta k = 2 k_{2} - k_{1} - k_{3} $. This wavevector mismatch  is naturally zero at the degenerate point where $\omega_1 = \omega_2 =\omega_3$, and it is  negative whenever $ \omega_3/\omega_1 >(n_2-n_1)/ (n_3-n_2)$, with $n_i = n(\omega_i)$  the refractive index. 
In most materials, the refractive index  decreases fast at shorter wavelengths, slower at longer wavelengths, and faster again at even longer wavelengths, where infrared oscillators have a progressively larger effect.
Because of this, when the pump wavelength is in the the shorter wavelength region,  $\Delta k$ initially becomes negative as $\omega_1$ decreases, moving away from the degenerate point, but then it becomes positive again as $\omega_1$ reaches the region where infrared oscillators start having a large impact  and $(n_3-n_2)< (n_2-n_1)$. It follows that for  ``short enough'' pump wavelengths there will always be a transition from $\Delta k <0$ to $\Delta k >0$ as $\omega_1$ decreases, and  phase matching can always be achieved. This point will be discussed at greater length later (Fig.~\ref{bulkAndSingleMode}).
 
In each of the three equations (\ref{cw1single}-\ref{cw3single}) the terms  related to self- and cross-phase modulation are in the first line, and the terms related to the third-order interaction between the three waves are in the second line. The third-order susceptibility $\chi_{\rm eff}^{(3)}$ that we defined in Eq.~(\ref{nlopol}) is characterized by the fact that it tends to the same off-resonant value in the limit where all its frequency arguments are small. For frequencies in the transparency region where the zero-frequency limit applies, it is therefore a good approximation to use the same $\chi_{\rm eff}^{(3)}$  for three-wave mixing and for cross- and self-phase modulation.

When these phase-modulation terms are neglected, one finds that the phase-matching condition $\Delta k=0$ will lead to a complete transfer of  the energy of the pump wave at frequency $\omega_2$ to the ``sidebands'' at frequencies $\omega_3$ and $\omega_1$ over an intensity-dependent conversion length. But this cannot work in general because the phase-modulation terms always lead to  dephasing of the waves with propagation, even when $\Delta k =0$. And while this dephasing is intensity-dependent and decreases at lower intensities, the same low intensity also leads to longer conversion lengths. In order to determine the conditions for optimum conversion efficiency it is therefore necessary to take  phase modulation into account. We  do this here by following an approach introduced by  Cappellini \emph{et al.} \cite{Cappellini91}.

We start by pointing out  that  in our case  the following quantities must be   independent from $z$ in the absence of  absorption:
\begin{eqnarray}
A_{tot}^2 &=& |A_{1}(z)|^2 + |A_2(z)|^2 + |A_{3}(z)|^2 \\
\Delta A^2 &=& |A_{3}(z)|^2 - |A_{1}(z)|^2
\end{eqnarray}
The first condition  reflects energy conservation, and $A_{tot}^2$ is the total photon flux carried by all three waves together. The second one is a characteristic of the nonlinear optical interaction we are discussing here, a process in which two photons are taken from wave 2 while one photon is added to each of the two other waves.

 In addition to $A_{tot}^2$ and   $\Delta A^2$, Cappellini \emph{et al.}  identified the following additional $z$-independent quantity (Eq.~6 in Ref.~\cite{Cappellini91}):
\begin{equation}
 |A_2(z)|^2 \left[4 |A_1(z)  A_3(z)| \cos \Delta \varphi(z)  +  \frac{\Delta k}{\chi  } + A_{tot}^{2}    - \tfrac{3}{2} |A_2(z)|^2 \right] \label{CappelliniInvariant}
\end{equation}
where $\Delta \varphi(z)= 2 \varphi_2 (z) - \varphi_1(z) - \varphi_3(z)$, and the $\varphi_i(z)$ are the phases of the complex amplitudes $A_i(z) = |A_i (z)| e^{i \varphi_i (z)}$ of the interacting waves.
 
We assume that the generated wave at frequency $\omega_1$ always starts at zero amplitude,   $A_1(0) = 0$. Eq.~(\ref{cw1single}) then implies that $\varphi_1(0) =  \pi/2 + 2 \varphi_2(0) - \varphi_3(0)$ and therefore $\Delta \varphi(0) =  2 \varphi_2(0) -\varphi_1(0)   - \varphi_3(0) = - \pi/2$. 

This value of the initial phase difference $\Delta \varphi(0)$ causes  the energy to flow from the pump wave to the two other waves. This situation would persist if there were no phase-modulation effects,  but the   self- and cross-phase modulation terms in Eqs.~(\ref{cw1single}-\ref{cw3single}) will always cause $\Delta \varphi(z)$ to drift away from this initial value: for $\Delta k=0$,
 the energy transfer from pump to generated waves cannot be maintained as the waves propagate.

Despite this, it is still possible to obtain a large conversion efficiency for a specific value of $\Delta k \neq 0$. To find this specific value, we use the fact that the maximum conversion efficiency, corresponding to a quantum yield of $\phi(z)=1$ (Eq. \ref{quantumyield})   is obtained when the pump wave has been completely depleted at some distance $z$. The condition $A_2(z) = 0$  then implies that  the invariant (\ref{CappelliniInvariant}) vanishes for all values of $z$, and in particular also for $z=0$, where $\cos \Delta \varphi(0)=0$. Inserting this in (\ref{CappelliniInvariant})  one finds that the optimum conversion efficiency can be realized when
$\Delta k = \chi \left[ (3/2) |A_2(0)|^2 - A_{tot}^2 \right]$, or
\begin{equation}
\Delta k = G =  \chi \left[\tfrac{1}{2} |A_2(0)|^2 - |A_3(0)|^2  \right] , \label{maxEffCondition}
\end{equation}
where we have defined an intensity-dependent phase modulation correction term $G$,  valid for $A_1(0) = 0$, which depends on the injected pump intensity (at $\omega_2$) and also takes into account the possible presence of a seed intensity (at $\omega_3$, in this case). The above condition corresponds to the conventional phase matching condition for low intensities, but it is otherwise a condition for optimum  conversion efficiency that depends only on the initial intensities and the third-order nonlinearity.

We  note that inserting Eqs.~(\ref{amplitudesDef}) and (\ref{chidef}) into   Eq.~(\ref{maxEffCondition}), together with the intensity $I = \epsilon_0 n c E^2/2$ and the nonlinear refractive index for self-phase modulation, $n_2^{NL} = 3  \chi^{(3)}_{\rm eff} / (4 \epsilon_0 c n^2)$, leads, in the limit of equal frequencies  and dominant pump intensity, to
\begin{equation}
G = \gamma P
\end{equation}
where $\gamma = 2 \pi n_2^{NL}/(\lambda A_{\rm eff})$ is the nonlinear coefficient of a fiber, with $A_{\rm eff}$  the effective mode area, and $P$  the average pump power in the guided mode. This result essentially matches earlier approximations in the limit of almost degenerate frequencies \cite{Yamamoto97,Song99,Serkland99,Wong07,Herzog12,Agrawal19}, confirming the validity of Eq.~(\ref{maxEffCondition}). 

The advantage of Eq.~(\ref{maxEffCondition}) compared to previous results is that it is an exact relationship, derived analytically in general for any three-wave interaction described by Eqs.~(\ref{cw1single}-\ref{cw2single}). As such, it is the necessary condition to obtain optimum quantum yield when taking into account both self-phase modulation and cross-phase modulation for all three interactive waves, and it will  be useful also when discussing the same effects in parallel, coupled waveguides in  section 3B. We also confirmed the validity of (\ref{maxEffCondition}) via direct numerical solutions of the coupled-wave equations.

As a rule of thumb, a gaussian laser pulse with an energy of 1 mJ, a 10 ns duration (full width at half maximum), and  a beam waist of 5.5~$\mu$m  has a peak intensity of $\sim 200$~GW cm$^{-2}$, which is close to the damage threshold of fused silica \cite{allison85, Smith09, Zervas14}. At such an intensity for the pump wave and no seed wave, the  phase modulation correction term $G$ in Eq.~(\ref{maxEffCondition})  has a value of  $\sim 100$~m$^{-1}$  (using $\chi^{(3)}_{1111} = 2 \times 10^{-22}$ m$^2$V$^{-2}$ for fused silica \cite{Milam98}). In fibers, we would expect at least one or two orders of magnitude less than this intensity.

\subsection{Wave interaction in two parallel waveguides}
For two parallel guiding cores, the wave interaction process is   described by six coupled-wave equations for the amplitudes of the waves  in the two cores $(a)$ and $(b)$. The three equations for core $(a)$ are the same as those for a single-core, but with the addition of the coupling terms to the other core:
\begin{eqnarray}
\dfrac{\partial}{\partial z}A^{(a)}_{1} & = & \textit{i} \: \chi [A^{(a)}_{1}(2 A^{(a)}_{3}A^{(a)^{*}}_{3} + 2 A^{(a)}_{2}A^{(a)^{*}}_{2} + A^{(a)}_{1}A^{(a)^{*}}_{1}) \nonumber \\ && + [A^{(a)}_{2}]^{2}A^{(a)^{*}}_{3} e^{\textit{i} \Delta k z}]   + \textit{i} \: \kappa_{1}A^{(b)}_{1},    \label{cw1} \\
\dfrac{\partial}{\partial z}A^{(a)}_{2} & =& \textit{i} \: \chi [A^{(a)}_{2} (2 A^{(a)}_{3}A^{(a)^{*}}_{3} + A^{(a)}_{2}A^{(a)^{*}}_{2} + 2 A^{(a)}_{1}A^{(a)^{*}}_{1}) \nonumber \\ && + 2 A^{(a)}_{1}A^{(a)^{*}}_{2}A^{(a)}_{3} e^{-\textit{i} \Delta k z}]   + \textit{i} \: \kappa_{2}A^{(b)}_{2},     \label{cw2} \\
\dfrac{\partial}{\partial z}A^{(a)}_{3} & =& \textit{i} \: \chi [A^{(a)}_{3}(A^{(a)}_{3}A^{(a)^{*}}_{3} + 2 A^{(a)}_{2}A^{(a)^{*}}_{2} + 2 A^{(a)}_{1}A^{(a)^{*}}_{1}) \nonumber \\ && + [A^{(a)}_{2}]^{2}A^{(a)^{*}}_{1} e^{\textit{i} \Delta k z}]  + \textit{i} \: \kappa_{3}A^{(b)}_{3}.  \label{cw3}
\end{eqnarray}
These equations are completed with  a symmetric set of  three equations describing the waves in core $(b)$. Here,  the $ \kappa_{i} = \kappa(\omega_{i}) $ are the wavelength-dependent coupling constants of each wave between the two cores; they can in general be expected  to be larger for longer wavelengths ($\kappa_1>\kappa_2>\kappa_3$). A wave of frequency $\omega_i$ launched in one waveguide will fully move to the other waveguide after a  length $L_c = \pi/(2 \kappa_i)$.
A similar set of equations has been analyzed in Refs.~\cite{Biaggio14,Ribeiro17}.  When it comes to just phase-matching, these equations can be  solved to obtain the phase matching conditions in  the limit where the phase modulation terms in (\ref{cw1}-\ref{cw3}) are neglected \cite{Biaggio14}. Doing so shows that phase matching is realized when any of the following conditions is fulfilled, \cite{Biaggio14}
\begin{eqnarray}
\label{CLPMrules1}  \Delta K_{\alpha} & = & \Delta k - (\kappa_{1}-\kappa_{3})  = 0,   \\
\label{CLPMrules2} \Delta  K_{\beta} & = &  \Delta k  - ( \kappa_{1} - 2\kappa_{2} +\kappa_{3} )  = 0,    \\
\label{CLPMrules3} \Delta  K_{\gamma} & = &   \Delta k  - (\kappa_{1} + 2\kappa_{2} +\kappa_{3}) = 0,  \\
\label{CLPMrules4}  \Delta K_{\alpha} ' & = & \Delta k + (\kappa_{1}-\kappa_{3})  = 0,    \\
\label{CLPMrules5} \Delta  K_{\beta} ' & = & \Delta k  + ( \kappa_{1} - 2\kappa_{2} +\kappa_{3} )  = 0,   \\
\label{CLPMrules6} \Delta  K_{\gamma} ' & = &  \Delta k  + (\kappa_{1} + 2\kappa_{2} +\kappa_{3}) = 0.  
\end{eqnarray}
These CLPM equations replace the simple $\Delta k = 0$ condition of the single-mode case by subtracting (as in Eqs.~\ref{CLPMrules1}-\ref{CLPMrules3}) or adding (as in Eqs.~\ref{CLPMrules4}-\ref{CLPMrules6})  specific linear combinations of the coupling constants. Thus, in this picture the coupling constants play the role of wavevectors that compensate for the phase mismatch described by $\Delta k$.

The coupling constants are related to the single-mode propagation constants $k_i$ and the propagation constants of the supermodes in the dual-core configuration by
\begin{equation}
  k_i = \frac{\beta^{(e)}_i + \beta^{(o)}_i}{2} , \ \ \  \kappa_i = \frac{\beta^{(e)}_i - \beta^{(o)}_i}{2} \label{kandkappas}
\end{equation} 
or
\begin{equation}
 \beta^{(e)}_i =  k_i+\kappa_i,  \ \ \ \beta^{(o)}_i = k_i-\kappa_i,  \label{betas}
\end{equation}
where  $\beta^{(e)}_i = n^{(e)}_i \omega_i/c$ and $\beta^{(o)}_i = n^{(o)}_i \omega_i/c$ are the propagation constants for the even and odd supermodes in the  two parallel waveguides, respectively. $n^{(e)}_i$ and $n^{(e)}_o$ are  the corresponding  refractive indices.

Substituting the relationships (\ref{kandkappas}) 
 into (\ref{CLPMrules1}-\ref{CLPMrules6}),  delivers the CLPM conditions in terms of the supermode propagation constants alone:
\begin{eqnarray}
\Delta K_\alpha&=& \beta^{(e)}_2 +\beta^{(o)}_2 -\beta^{(e)}_1  - \beta^{(o)}_3 = 0,  \label{bt1}\\
\Delta K_\beta &=& 2 \beta^{(e)}_2 -\beta^{(e)}_1  - \beta^{(e)}_3= 0,  \label{bt2}\\
\Delta K_\gamma &=&   2 \beta^{(o)}_2 -\beta^{(e)}_1   - \beta^{(e)}_3 = 0,  \label{bt3} \\
\Delta K_\alpha ' &=&   \beta^{(e)}_2 + \beta^{(o)}_2 -\beta^{(o)}_1  - \beta^{(e)}_3 = 0,  \label{bt4}\\
\Delta K_\beta '&=& 2 \beta^{(o)}_2  -\beta^{(o)}_1 - \beta^{(o)}_3 = 0,  \label{bt5}\\
\Delta K_\gamma '  &=&  2 \beta^{(e)}_2 -\beta^{(o)}_1  - \beta^{(o)}_3 = 0  . \label{bt6}
\end{eqnarray}
CLPM phase matching, which was initially expressed in terms of the coupling constants alone \cite{Biaggio14}, is thus clearly revealed to correspond to phase-matched mode-coupling between the supermodes of the dual core structure. Both pictures are equally valid and can give complementary insights. 

By moving to a dual core configuration it becomes possible to choose between six different phase matching conditions instead of only one for a single-mode fiber. The limited  phase matching condition for   the single-mode case, which  can be adjusted  by varying core-radius and refractive index contrast, is greatly expanded by  the  possibility of varying  the distance between the two cores. This additional flexibility of CLPM in dual core fibers also greatly expands the range of available pump wavelengths and generated wavelengths.

\begin{figure}[t!]
  		 \centerline{\includegraphics[width=\linewidth]{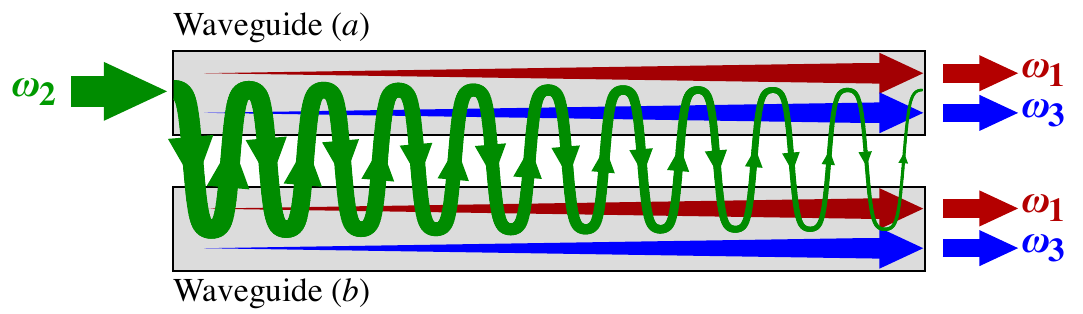} }
\caption{Cartoon of the CLPM idea for  the $2 \omega_2 \to \omega_1 + \omega_3$ third-order three-wave mixing process in dual-core fibers. A pump wave (green) is injected into one of two parallel waveguides, causing it to zig-zag between the two waveguides. Then, because of third-order optical nonlinearities, two additional waves, one at a shorter wavelength (blue), and one at a longer wavelength (red), are created during propagation. By adjusting the distance between the waveguide, one can obtain phase-matched energy transfer between the three waves  for almost any choice of wavelengths. This picture approximates the situation when the $2 \omega_2 \to \omega_1 + \omega_3$ process is phase matched via the $\Delta K_\alpha$ CLPM condition, in which potentially the full power of the pump wave can be  converted into  two generated waves that, over long distances,  mostly grow in  the even or odd supermode of the two-waveguide structure. } \label{introfig}
\end{figure}

To conclude this section, we present in Fig.~\ref{introfig} a simplified pictorial summary of the energy transfer from a pump wave to generated waves that can be obtained thanks to CLPM conditions such as $\Delta K_\alpha =0$ (Eq.~ \ref{CLPMrules1} or \ref{bt1}) in two parallel waveguides. The pump wave is initially injected in just one of the two coupled waveguides, and its power will then move back and forth between the two waveguides during propagation. Concurrently, the third-order nonlinear interaction transfers energy from the pump wave to the generated waves. Under the appropriate conditions, this energy transfer continues unabated as the waves propagate, leading to the depletion of the pump wave and the creation of the generated waves. This can be obtained via the CLPM conditions and by taking into account self-and cross-phase modulation between all waves. The way this works can be understood in this case in a manner similar to the quasi-phase-matching process in second order nonlinear optics. In one of the two waveguides, the intensity of the pump wave is spatially modulated by the coupling to the other waveguide. The creation of the generated waves therefore also happens in the same spatially modulated way. This then allows the phase of the waves to be the affected by the three-wave interaction when the pump wave is present, and  by linear optical propagation when it is not there. When the combination of the two phase changes is ``just right'', phase matching can be obtained. However, words are not enough to describe what happens, and the situation is significantly more complicated because the coupling to the other waveguide actually affects all waves, and because self- and cross-phase modulation also influence the phase of the waves as they propagate.

The influence of cross- and self-phase modulation on the quantum yield of the three-wave interaction process in dual-core fibers, and the dual-core equivalent of the phase-modulation correction term in Eq.~(\ref{maxEffCondition}), will be discussed in the next section (part B), which also provides several examples of how the CLPM conditions can be realized by adjusting the fiber design parameters.

\section{Discussion of frequency down-conversion in step-index fibers}
The following discusses  phase-matched three-wave mixing in single and dual-core fibers, and describes the conditions for achieving optimum conversion efficiency in the presence of self- and cross-phase modulation. 

The topic and relevant issues are first highlighted for simple plane-wave interactions and for a single   cylindrical step-index waveguide in the example of a single-mode fiber. This will also serve as a review of past work and as an introduction to the topic, before going on to the dual-core, or twin-core, system and the advantages of employing coupling-length phase matching there.

Most of the calculations presented below compute CLPM conditions  using Eq.~(\ref{kappaEq}) and Eqs.~(\ref{CLPMrules1}-\ref{CLPMrules6}). In some cases the accuracy of the calculation was checked using COMSOL Multiphysics \cite{COMSOL} to find the propagation constants of the supermodes, and then either   Eq.~(\ref{kandkappas}) to obtain the single-mode propagation constant, effective refractive index, and coupling constants, or else  directly the equivalent expressions  (\ref{bt1}-\ref{bt6}). 

Since calculations based on  Eq.~(\ref{kappaEq}) loose any significance for a wave with a frequency approaching the cut-off frequency of the odd supermode, we also used COMSOL  \cite{COMSOL} to check for such cases, or we generally avoided this situation by focusing on the CLPM conditions (\ref{bt1}-\ref{bt3}), which do not rely on the odd supermode for the generated wave at the longest wavelength. Another reason to favor generation in the even supermode is the fact that it facilitates outcoupling of the generated wave.

\subsection{Free space propagation and single-mode waveguides}

Since the coupled-wave equations (\ref{cw1single}-\ref{cw3single}) are valid both for a single-mode fiber as well as for plane-wave propagation in the bulk material, it is useful to first consider the behavior of the phase-mismatch  $\Delta k$ as determined by the refractive index dispersion in typical glasses. In contrast to simple second-order nonlinear optical interactions, the fact that the refractive index monotonically decreases does not forbid phase matching for the $2 \omega_2   \to \omega_1 + \omega_3$ process (which is also naturally phase-matched in the degenerate case where all frequencies are equal). 

\begin{figure}[ht!]
  		 \centerline{\includegraphics[width=0.92\linewidth]{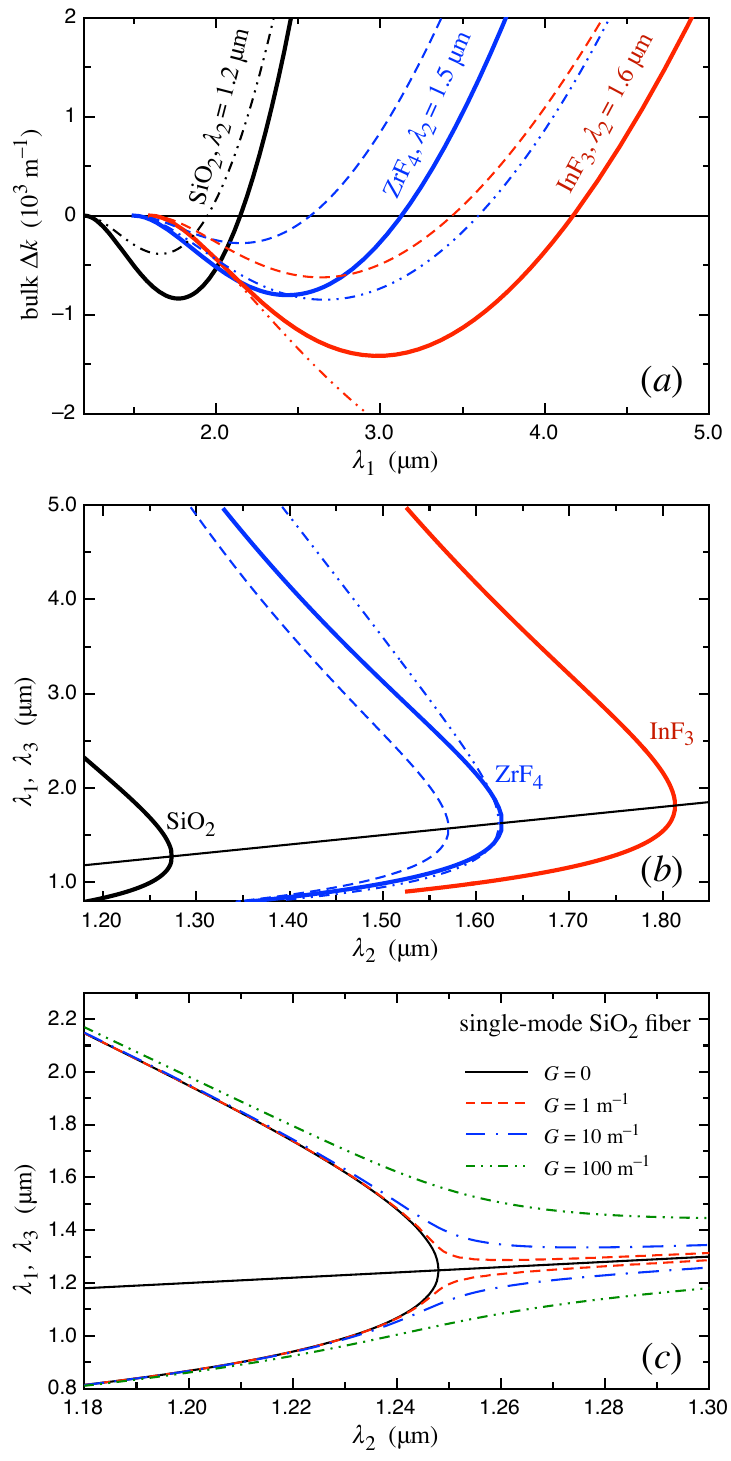} }
\caption{Frequency conversion for  plane-wave interaction and for single-mode fibers. Top: Phase mismatch  $\Delta k$ vs.~generated wavelength $\lambda_1$,  for different pump wavelengths and glass types. The differently colored curves are labeled with the corresponding material. Solid curves are for plane waves,  dashed and dashed-dotted curves of the same color are  for single-mode step-index fibers with an index contrast of $\Delta n =0.01$ and a core radius of 9 $\mu$m and 4.5 $\mu$m, respectively. 
Middle: Wavelength-tuning curves for the generated wavelengths as a function of the pump wavelength for the three glasses. Thick solid curves are for plane-wave interaction.  The solid line is for the degenerate case of all frequencies equal to each other. The ZrF$_4$ example includes curves  for a 4.5$~\mu$m core fiber (dash-dotted line) and a 9$~\mu$m core fiber (dashed line).
Bottom: Wavelengths for optimum single-mode frequency downconversion efficiency in a silica fiber ($\Delta n = 0.01$, $r = 4.5 \ \mu$m) when neglecting self- and cross-phase modulation ($G=0$), and for various  phase modulation correction terms $G$   (Eq.~\ref{maxEffCondition}).} \label{bulkAndSingleMode}
\end{figure}

A plot of $\Delta k(\lambda_1)$ for different  pump wavelengths $\lambda_2$ is shown in Fig.~\ref{bulkAndSingleMode}($a$). The phase-matching conditions, given by the wavelengths at which the curves cross the $\Delta k =0$ line, change when changing pump wavelength or, in fibers,  also when changing refractive index contrast or core radius. To get a better overall view of the phase-matching possibilities, it is more useful to plot the generated wavelengths as a function of the pump wavelength  for given geometrical parameters: this is done for the same three glasses in  Fig.~\ref{bulkAndSingleMode}($b$) for plane-wave interaction. In order not to overload these examples, curves that apply to fibers are only included   for  the example of ZrF$_4$. In these plots, each phase-matching curve reaches a cusp at a maximum pump wavelength above which no phase matching can be obtained. The cusp of the general wavelength-tuning curves corresponds to the wavelength of  zero group-velocity dispersion, which is also marked by the intersection with the ``degeneracy line'' for the case of three equal frequencies.   For pump wavelengths  below this maximum value, the two branches of the curves give the wavelength of the generated waves, which move  farther and farther apart as the pump wavelength decreases \cite{garth1986,sammut1989,Agrawal19book}.  

Here and in the following, all our examples will be for step-index cylindrical waveguides with an index contrast of $\Delta n = 0.01$, and most of them will use a core radius of $4.5 ~\mu$m. As a general observation, a silica waveguide with these properties has a zero group-velocity dispersion wavelength shorter than for the corresponding bulk glass, which leads to  the phase-matching curves of Fig.~\ref{bulkAndSingleMode}($b$) being shifted farther to the left when going from plane-wave interaction to  waveguides with these parameters. It would take up too much space in this contribution to consider all possible values of radius and index contrast, and we concentrate instead on examples that clarify the most important features. Suffice it to say that a decrease in core radius or a decrease in refractive index contrast around the chosen values have a similar effect, shifting phase-matching curves like those in Fig.~\ref{bulkAndSingleMode}($b$) to the right, towards larger pump wavelengths (this is seen for the ZrF$_4$ waveguide, which for a $9 \ \mu$m core radius has the phase matching curves shifted to lower pump wavelengths compared to the bulk, but then the cusp of the curve is shifted again towards longer pump wavelengths when the radius is decreased to $4.5 \ \mu$m).

From Fig.~\ref{bulkAndSingleMode}($b$), one finds that   the $2 \omega_2   \to \omega_1 + \omega_3$ process  can  generate wavelengths between 2.5 and 1.7 $\mu$m in bulk silica when using pump wavelengths between $\sim 1.16$ and $\sim 1.25$~$\mu$m, respectively. And in bulk ZrF$_4$ (or InF$_3$) glass the same process can generate wavelengths between 5 and 2.5 $\mu$m when using pump wavelengths between $\sim 1.33$ and $\sim 1.57$ ($\sim 1.53$ and $\sim 1.77$ for InF$_3$) $\mu$m. In waveguides made with the same glasses, waveguide dispersion modifies the guided-wave $\Delta k$ but the main characteristics of the phase-matching condition remain similar. The possibility of obtaining $\Delta k = 0$ for three-wave mixing  was recognized early  on for  single-mode fibers \cite{Lin81,Lin82,Lin83}, but it is a general feature  that   applies  also to plane-wave interaction in all bulk materials with normal dispersion (see also section 2\ref{singlemodeinteraction}). Despite this, we are not aware of any experimental demonstration of the predictions of  Fig.~\ref{bulkAndSingleMode}($b$)  in  bulk glasses.

Fig.~\ref{bulkAndSingleMode}($c$) shows how the wavelength-tuning curves of a single-mode silica fiber mutate from giving just the phase matching condition in the low-intensity limit to giving the condition for  optimum conversion efficiency  when self- and cross-phase modulation come into play. This plot gives the   wavelength choices, determined from  Eq.~(\ref{maxEffCondition}), that allow to reach optimum quantum yield (\ref{quantumyield}) for a long enough interaction length. 
For low intensities ($G \to 0$) all curves in Fig.~\ref{bulkAndSingleMode}($c$) tend to the standard phase-matching curve, with the  maximum pump wavelength at $\lambda_2^{max} = 1.248$~$\mu$m. 
For larger values of $G$,  the generated wavelength $\lambda_1$ increases, an effect that is stronger near the degeneracy point at $\lambda_2^{max} = 1.248$~$\mu$m. This result implies a slight intensity-dependent shift in the sidebands generated by a single pump wave   \cite{Harvey03,Wong07}. While this also leads to the possibility of achieving efficient interactions for pump wavelengths above  $\lambda_2^{max}$, the generated wavelengths are limited to a small interval close to the pump wavelength. We will build upon this result in the next section when analyzing a dual core configuration, which will allow to obtain a phase-matched interaction over a much wider range of generated wavelengths at larger pump wavelengths.

\begin{figure}[hbt!]
  		 \centerline{\includegraphics[width=0.95\linewidth]{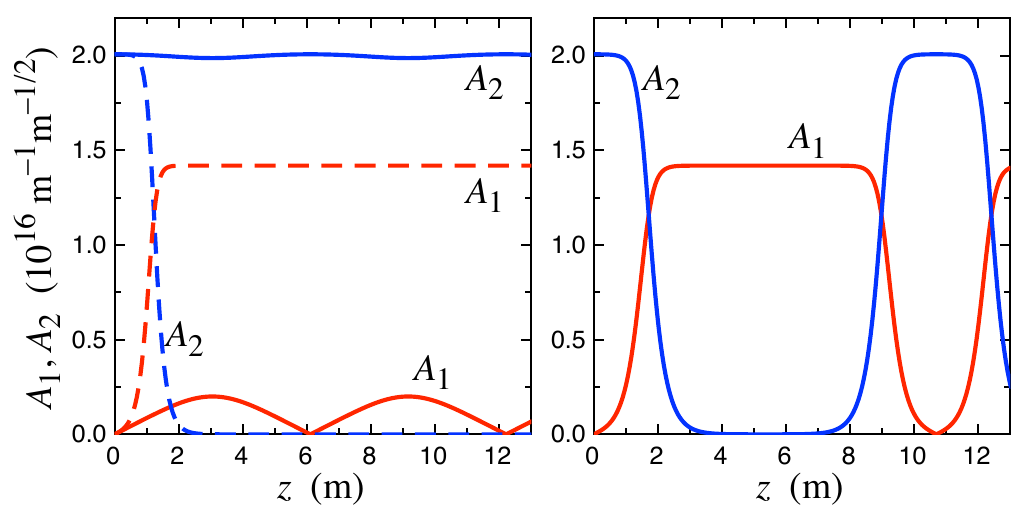} }
\caption{Solution of the coupled-wave equations (\ref{cw1single}-\ref{cw3single})  in a single-mode silica fiber ($\Delta n = 0.01, r=4.5 \ \mu{\rm m}$). Left: $\Delta k = 0$ in the presence of phase-modulation (solid curves) and without taking into account phase modulation (dashed curves). Right: Solution when fully taking into account phase modulation by setting $\Delta k = G$. For this example, $A_2(0)=2 \times 10^{16}$ m$^{-1}$s$^{-1/2}$, $G = 2.15 \ {\rm m}^{-1}$, and $\chi = 1.04 \times 10^{-32}$~m (as obtained using (\ref{chidef}) and the third-order susceptibility of fused silica of $2 \times 10^{-22} \ {\rm m}^2 {\rm V}^{-2}$). The pump wavelength is kept at $\lambda_2 = 1.2 \ \mu$m. The generated wavelength corresponding to the $\Delta k =0$ and $\Delta k = G$ conditions is $\lambda_1=1.9480 \ \mu$m and $\lambda_1=1.9493  \ \mu$m, respectively.} \label{singlemodesilicapropagation}
\end{figure}

To further illustrate the effect of phase-modulation on the conversion efficiency, Fig.~\ref{singlemodesilicapropagation} shows a numerical solution of the coupled-wave equations (\ref{cw1single}-\ref{cw3single}) for   intensities such that $G=2.15$~m$^{-1}$ and for two choices of wavelengths: one that satisfies  $\Delta k =0$ and one that satisfies $\Delta k = G$.  $\Delta k =0$ gives a 100\% conversion efficiency  when neglecting  phase modulation (dashed curves), but gives only  little frequency conversion otherwise (left graph). On the other hand, 100\% conversion efficiency is recovered when the wavelengths of the interacting waves are tuned in such a way that $\Delta k = G$ (right graph). We note that the figure plots amplitudes, and therefore conversion efficiency approaching 100\% is obtained at a propagation distance $z$ where  $A_1(z) \to  A_2(0)/\sqrt{2}$.

The initial conditions chosen for Fig.~\ref{singlemodesilicapropagation} are $A_1(0)=0$ and $A_2(0) = 100 A_3(0)$. This  effectively approximates, using the coupled-wave equations (\ref{cw1single}-\ref{cw3single}), a situation where the fiber is only injected with the pump wave at frequency $\omega_2$, with the other two waves arising from noise. In such a situation, the growth of the generated wave amplitude as it approaches maximum conversion efficiency deviates from the familiar   linear growth of $A_1(z)$ that then saturates towards optimum conversion. In addition, it is worth mentioning that in the case when the fiber were only pumped by a single wave, the system would spontaneously choose the generated wavelengths that fulfill $\Delta k = G$. The effect of phase modulation leads to a slight  (but predictable) increase in  the generated wavelength $\lambda_1$ as the intensity of the pump wave increases.

For the calculations in Fig.~\ref{singlemodesilicapropagation}, we assumed the availability of nanosecond duration pulses with energies of the order of microjoules, which have been demonstrated for various fiber laser systems \cite{Pavlov14, Lee18}. The intensity corresponding to the initial amplitude of the pump wave used for Fig.~\ref{singlemodesilicapropagation} is equal to 
$3.4~\times~10^{13} { \rm \ W m}^{-2} = 34 {\rm \ W} \mu {\rm m}^{-2}$, which is the peak intensity of a 10 ns pulse with an energy of 10 $\mu$J when focused to a beam waist of 4.5 $\mu$m. Choosing other values for the pump wave intensity leads to different propagation lengths for optimum frequency conversion, but it does not otherwise change the behavior of the solutions (the conversion length is in general inversely proportional to the pump intensity).   It is necessary to choose pulse durations of several nanoseconds or above in order to avoid   the walk-off of the interacting pulses caused by group-velocity dispersion. For the case of frequency downconversion from 1.2-1.5 $\mu$m wavelengths to $2 ~\mu$m wavelength  in silica fibers, the interaction distance for pulse durations below  1 ns pulses would be limited to less than a few meters, decreasing linearly with the pulse duration.

\subsection{Dual core fibers}

\begin{figure}[ht!]
  		 \centerline{\includegraphics[width=\linewidth]{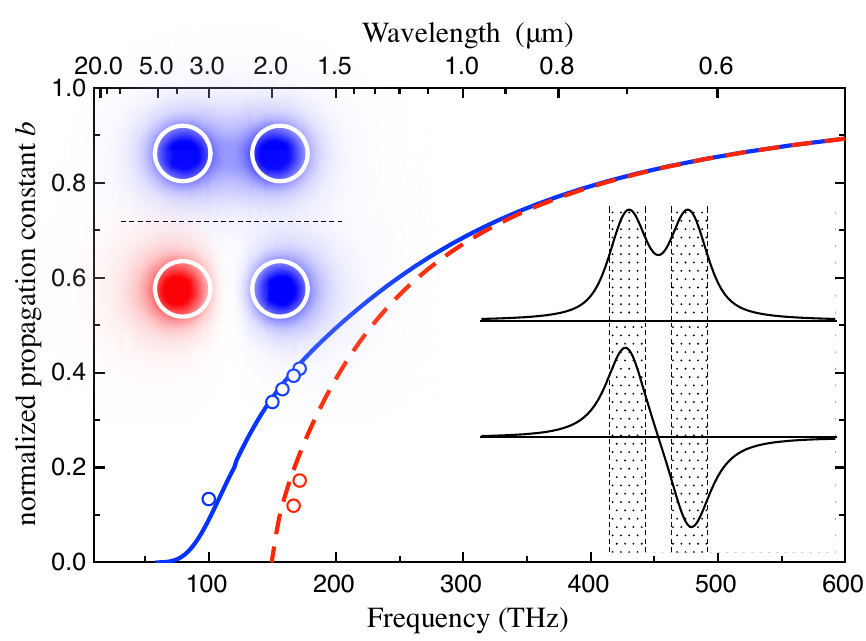} }
\caption{Dispersion of the even (solid curve) and odd (dashed curve) supermodes in a step-index silica fiber with a core radius of 2 $\mu$m, a core distance of $7 \ \mu$m, and $\Delta n =0.01$. The curves for the normalized propagation constants are obtained from Eqs.~(\ref{kappaEq}), (\ref{betas}), and (\ref{nEff}). The data points close to the curves represent the corresponding data obtained in COMSOL. The insets shows the  amplitude profiles of the two supermodes in the dual core structure of this example, at  a wavelength of 1.55 $\mu$m,  as plots of the electric field amplitude along a line connecting the cores in the bottom-right corner, and as color plots in the top-left corner (where the boundary  between each cylindrical core and the cladding is given by the thick white line).} \label{modedispersiondual}
\end{figure}

Dual core fibers can be understood  in terms of the modes propagating in one of the two cores and how they couple to the modes in the other core, which lead to the CLPM expressions (\ref{CLPMrules1}-\ref{CLPMrules6}), or they can also be understood in terms of the two lowest frequency supermodes of the dual-core structure, which lead to the alternative form of the CLPM expressions, (\ref{bt1}-\ref{bt6}). Fig.~\ref{modedispersiondual} shows an example of the profile and dispersion of the two lowest-frequency supermodes of two parallel cylindrical cores.  One main issue that will be addressed below is that at large wavelengths and small distances between the cores it is possible that the odd supermode is not allowed anymore. This must  be taken  into account when working with expressions such as (\ref{CLPMrules1}-\ref{CLPMrules6}).

The design parameters that come into question in order to realize the CLPM conditions are the radius of each cylindrical core, the distance of the cores, and the refractive index contrast. In the following we will discuss examples for the case of simple step-index  fibers with a difference between cladding and core refractive index of 0.01, focusing on how the CLPM conditions can be tuned by the wavelengths participating in the interaction, and by geometrical parameters, in particular the distance between the cores.

\subsubsection{Silica dual core fibers}
\begin{figure}[ht!]
  		 \centerline{\includegraphics[width=\linewidth]{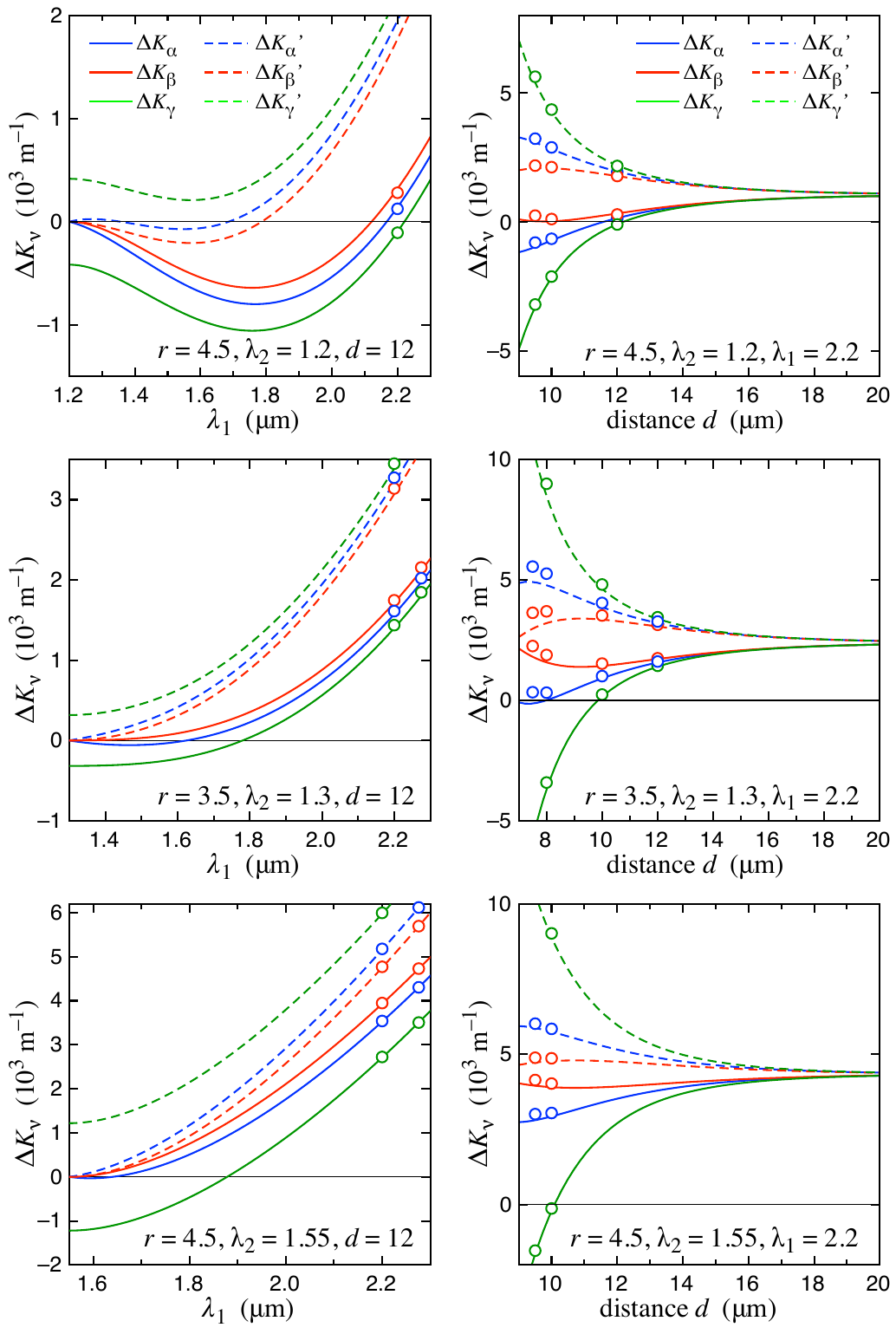} }
\caption{Phase mis-match terms $\Delta K_\nu$ (From Eqs.~\ref{CLPMrules1}-\ref{CLPMrules6}) for a dual core step-index silica fiber, as a function of the generated wavelength $\lambda_1$ (plots on the left, for $d = 12$~$\mu$m), and as a function of the  center-to-center distance $d$ between the cores  (plots on the right, for $\lambda_1=2.2$~$\mu$m).  All the relevant parameters are given in the individual plots. For the pump wavelength of $\lambda_2=1.3$~$\mu$m we chose a smaller radius $r=3.5$ $\mu$m in order to allow phase matching using the $\Delta K_\alpha=0$ condition (see text).  The order of the curves in the plots, from top to bottom, corresponds to $\Delta K_\gamma '$, $\Delta K_\alpha '$, $\Delta K_\beta '$, $\Delta K_\beta $, $\Delta K_\alpha$, $\Delta K_\gamma $. The data points on top of some curves were obtained from COMSOL and Eqs.~(\ref{bt1}-\ref{bt6}), to assess the accuracy of these calculations.} \label{AllCLPMconds}
\end{figure}

The six expressions for the phase-mismatch terms $\Delta K_\nu$ for frequency down-conversion in parallel waveguides, given in complementary ways  in Eqs.~(\ref{CLPMrules1}-\ref{CLPMrules6}) or  in Eqs.~(\ref{bt1}-\ref{bt6}), are plotted for a dual-core step-index silica fiber in Fig.~\ref{AllCLPMconds} as a function of the generated wavelength $\lambda_1$ and the distance between the cores, and  for different choices of core radius and   pump wavelength $\lambda_2$. 

The dashed curves in the figure belong to $\Delta K_\alpha '$, $\Delta K_\beta '$, and $\Delta K_\gamma '$, for which the linear combinations of coupling constants is added to $\Delta k$. They can only cross zero (satisfying the CLPM condition) when the pump wavelength is short enough (top-left panel in the figure), and the corresponding phase-matching wavelength $\lambda_1$ is shorter than that obtained in a single-core fiber for the same core radius (see Fig.~\ref{bulkAndSingleMode}). On the other hand, the solid curves belonging to $\Delta K_\alpha$, $\Delta K_\beta$, and $\Delta K_\gamma$, where the coupling constants are subtracted from the single-mode $\Delta k$,  generally cross zero at wavelengths larger than the phase-matching wavelength $\lambda_1$ for a single-core fiber.

Another  important difference between these two families of CLPM conditions is the fact that the three conditions corresponding to solid curves in  Fig.~\ref{AllCLPMconds} cause the generated wavelength to appear in the even supermode of the dual-core structure, as can be clearly seen from Eqs.~(\ref{bt1}-\ref{bt3}), while the three drawn with dashed cruves let the generated wavelength $\lambda_1$ appear in the odd supermode (see Eqs.~\ref{bt4}-\ref{bt6}). The latter case would make it more difficult to extract the generated wave from the fiber, but the fact that this odd mode is required for phase matching also eliminates the possibility of using the corresponding phase matching conditions  when the generated frequency is below the cut-off frequency for the odd supermode. This is shown quantitatively in the standard mode-dispersion plot of Fig.~\ref{modedispersiondual}. 

The CLPM condition $\Delta K_\beta=0$ applies when all interacting waves propagate in the lowest-frequency fundamental mode (the even supermode), and its possible realization depends on  characteristics of  the bulk refractive index dispersion, as shown in the previous section. In fact, the quantity $\Delta K_\beta$ of Eq.~(\ref{bt2}) directly corresponds to the standard phase-mismatch for the lowest-frequency mode, and it is essentially the equivalent of the quantity plotted in Fig.~\ref{bulkAndSingleMode}, with a weak dependence from the distance between the cores.

From Eqs.~(\ref{bt2}) and (\ref{bt3}) one also notices that the corresponding CLPM conditions can only maximize the conversion  efficiency when the pump wave is launched into exactly one supermode, either even or odd. When the pump wave is launched in only one of the two cores (and is therefore  a superposition of even and odd supermode),  the conversion efficiency is limited to a maximum theoretical efficiency  50\%,  because only one of the supermodes excited in this way benefits from phase-matched coupling to the generated wave, and therefore only half of the injected pump-wave photons can be used to create the generate wave  \cite{Biaggio14}. The only way to obtain a  larger theoretical conversion efficiency using $\Delta K_\gamma$, Eq.~(\ref{bt3}), would be to inject the pump wave at $\omega_2$ as an odd-supermode, which seems like an unnecessary complication for practical applications. In addition, there are practical reasons why the theoretical efficiency cannot be reached, and it is possible that a practical realization will prefer to use CLPM condition (\ref{bt3}) even though its theoretical efficiency is lower.

In the following we will concentrate on the case where the pump wave at $\omega_2$ is injected in one core only, which then leads to its intensity oscillating between the two cores, and on the use of  the CLPM conditions $\Delta K_\alpha = 0$ and $\Delta K_\gamma = 0$ (Eqs \ref{CLPMrules1} and \ref{CLPMrules3}, or Eqs.~\ref{bt1} and \ref{bt3}). 

The first of these two CLPM conditions is the best one to achieve optimum conversion efficiency when the pump wave is injected in just one core because it requires the presence of both the odd and even supermodes for the pump wave. Its other advantages are that it does not depend on the coupling constant $\kappa_2$ and that (as we will see later) the corresponding nonlinear interaction is not affected by  self- and cross-phase modulation.

The second of these two CLPM conditions, $\Delta K_\gamma =  0$, is also attractive(\ref{bt3}), despite the fact that its optimum conversion efficiency is limited   50\%,  because it  extends the range of pump wavelengths over which CLPM can be obtained.

\begin{figure}[h!]
  		 \centerline{\includegraphics[width=\linewidth]{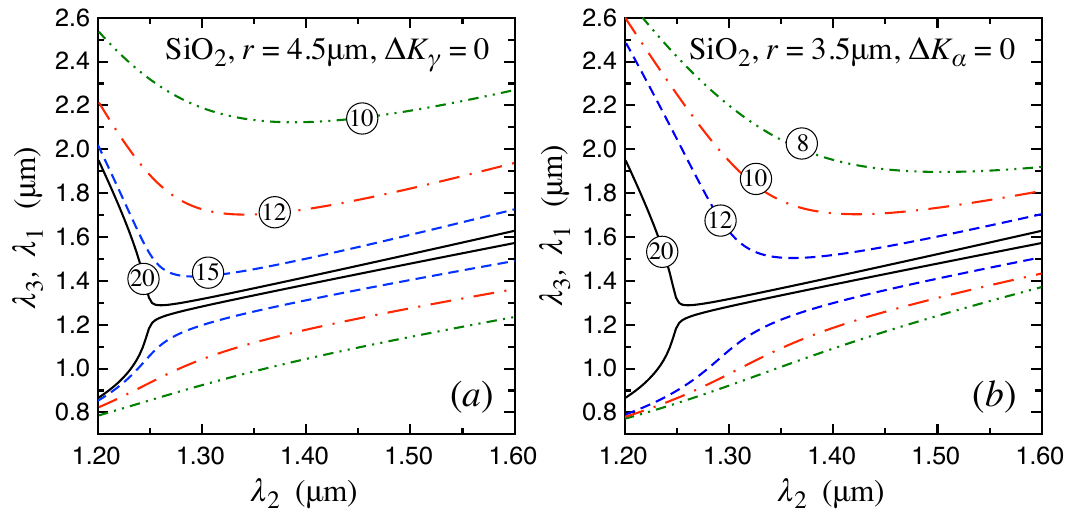} }
\caption{Wavelength-tuning CLPM curves  for different distances between the cores in a dual core step-index fiber with a refractive index contrast of $\Delta n=0.01$. Left: $\Delta K_\gamma=0$ for a core radius of 4.5 $\mu$m. Right: $\Delta K_\alpha=0$ for a core radius of 3.5 $\mu$m. The wavelength-tuning curves are labeled with the center-to-center distance between the cores in micrometers.} \label{CLPMwlDistance}
\end{figure}

\begin{figure}[h!]
  		 \centerline{\includegraphics[width=\linewidth]{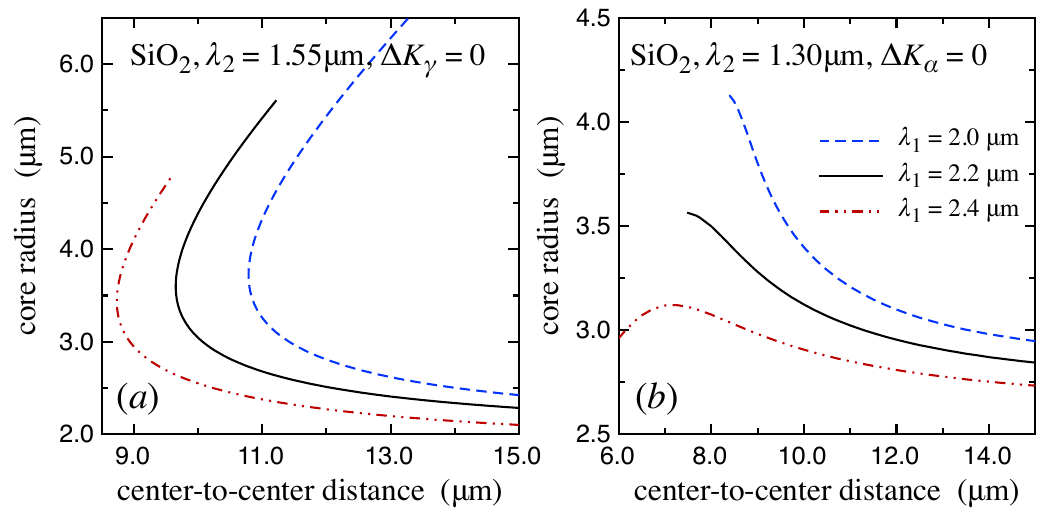} }
\caption{Geometry-tuning CLPM curves for different down-converted wavelengths  in a dual core step-index fiber with a refractive index contrast of $\Delta n=0.01$. Left: $\Delta K_\gamma=0$ for a pump wavelengths  $\lambda_2 = 1.55$~$\mu$m. Right: $\Delta K_\alpha=0$ for a pump wavelength  $\lambda_2 = 1.30$~$\mu$m.} \label{CLPMrDwavelengths}
\end{figure}

Fig.~\ref{CLPMwlDistance} shows wavelength-tuning curves similar to those plotted in Fig.~\ref{bulkAndSingleMode}($b$) for different center-to-center distance between the two cores. The  phase matching curve in the limit of large distances between the cores corresponds to that plotted in Fig.~\ref{bulkAndSingleMode}($b$). But as the distance between the cores decreases, the coupling between the cores  extends the phase matching possibilities to a much wider range of pump wavelengths. This is because the effect of the coupling as the distance between the cores decreases  adds a distance-dependent offset to the condition $\Delta k = 0$, as seen in Eqs.~(\ref{CLPMrules1}-\ref{CLPMrules6}), which changes the shape of the wavelength-tuning CLPM curves in a slightly different way for the two conditions ($\Delta K_\alpha=0$ and $\Delta K_\gamma=0$) that we are considering here. The way this happens also depends on the radius of the core, which  influences the coupling. For $\Delta K_\alpha=0$, the generated wavelength $\lambda_1$ moves to smaller values as the pump wavelength increases, but it can be maintained above 2 $\mu$m by going to smaller pump wavelengths $\lambda_2$ while also reducing the core-radius. Because of this, we chose to plot the wavelength-tuning curves for this condition using a core radius of $3.5$~$\mu$m.
	
Fig.~\ref{CLPMrDwavelengths} shows geometry-tuning curves describing how the same two CLPM conditions vary with core radius and distance  for a fixed choice of interacting wavelengths.  The plot for $\Delta K_\alpha=0$ is for a pump wavelength of $1.3$~$\mu$m, while we chose a pump wavelength of $1.55$~$\mu$m for  $\Delta K_\gamma=0$. For both cases we considered the same generated wavelengths between $\lambda_1=2.0$~$\mu $m and $\lambda_1=2.4$~$\mu$m. The plot for the pump-wavelength $\lambda_2=1.3$~$\mu$m shows again how it is necessary to reduce the core-radius in order to achieve generation of wavelengths of the order of $\lambda_1=2.2$~$\mu$m. 
The choice of pump wavelengths for the  above examples  corresponds to those available from fiber lasers such as those based on praseodymium  for the 1.3 $\mu$m pump wavelength, and Erbium for  the 1.55 $\mu$m wavelength.

Next, we address the effect of self-and cross-phase modulation on the three-wave mixing process and the corresponding ability to achieve  maximum conversion efficiency.
The effect of phase modulation will also depend on  how the pump wave is injected into the system and on the specific CLPM condition, making an analytical approach cumbersome. As an example, Ref.~\cite{Ribeiro17} provides an analysis of a similar dual core system but ignores the frequency dependence of the coupling constants and the six different CLPM conditions. An analytical approach similar to that we successfully employed  for a single-mode fiber is not as straightforwardly applicable to  the dual-core system. On the other hand, a   numerical solution of the coupled-wave equations (\ref{cw1}-\ref{cw3}) can be easily implemented and can be effectively used to assess the effect of phase modulation on the dependence of the wave amplitudes on propagation distance under all different injection schemes, pump intensities, and  CLPM conditions. In this way one arrives at a quantification of the effect of  phase-modulation on the quantum yield for an optimized interaction length. 

The resulting phase-modulation correction factors, and the dual-core equivalent to   Eq.~\ref{maxEffCondition} for our preferred CLPM conditions discussed above, are 
\begin{eqnarray}
	\Delta K_\alpha &=& 0  \label{pmCorrDualalpha} \\
	\label{corrPMKgamma} \Delta K_\gamma &=& \frac{G}{4} + \eta   G^2 , \label{pmCorrDualgamma}
\end{eqnarray}
where  $\eta$ is a small correction ($\eta  G \ll 1$) that is essentially intensity-independent, and can change slightly depending on the coupling between the cores. But for any practical application it does not significantly affect the condition \ref{pmCorrDualgamma} and the CLPM conditions derived from it.

\begin{figure}[h!]
  		 \centerline{\includegraphics[width=\linewidth]{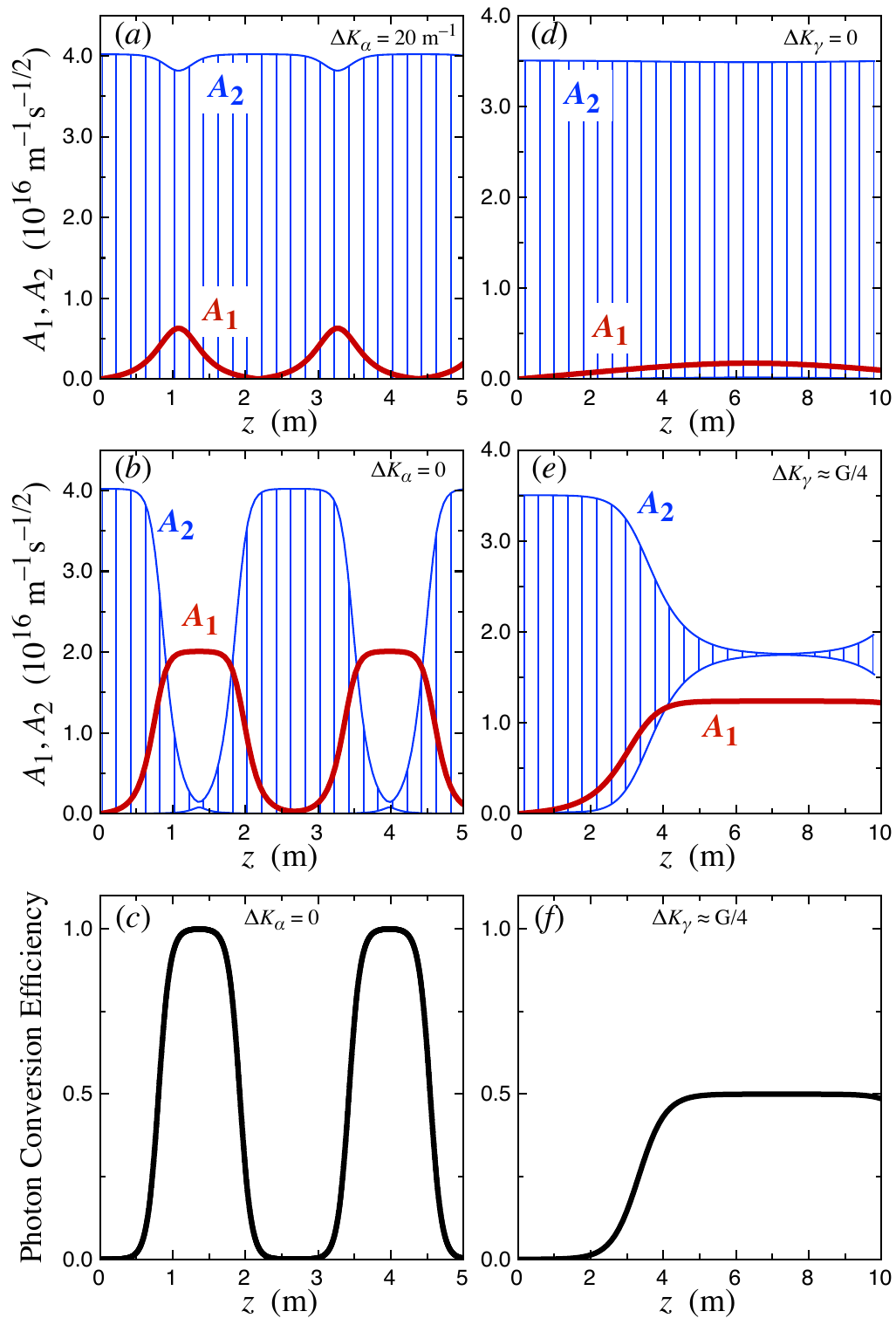} }
\caption{Amplitudes of the interacting waves  and photon conversion efficiency (quantum yield $\phi$ from Eq.~\ref{quantumyield})   vs. propagation distance for a dual-core silica fiber with refractive index contrast of 0.01 in CLPM configurations that lead to frequency downconversion to a generated wave with $\lambda_1=2.2$~$\mu$m. While the curves for $A_1$ are the full $z$-dependence, showing that the generated wave does not oscillate between the fiber cores, the curves for $A_2$ are the \emph{envelope} of  oscillations that occur over coupling lengths of the order of millimeters, the vertical stripes are drawn between the minimum and the maximum of the oscillations.   The calculations assume a 10 ns long pump-wave pulse with an energy of 25 $\mu$J with optimum coupling into one single core (peak intensity $\sim 10^{14}$ W/m$^2$).
\\
Graphs in the left column ($a$-$c$):  CLPM configuration $\Delta K_\alpha$. Pump wave at $\lambda_2=1.3$~$\mu$m, core radius $r=3.5$~$\mu$m, center-to-center core distance $d=7.99$~$\mu$m, G=7.4~m$^{-1}$.  $\Delta K_\alpha=0$ is not fulfilled for  ($a$), and is  fulfilled for  ($b$) and ($c$).
\\
Graphs in the right column ($d$-$f$):  CLPM configuration $\Delta K_\gamma$. Pump wave at $\lambda_2=1.55$~$\mu$m, core radius $r=4.5$~$\mu$m,  $d=10.09$~$\mu$m,  G=4.1~m$^{-1}$. Panel ($d$) is for $\Delta K_\gamma=0$  but  without the phase-modulation correction, while for panels ($e$) and  ($f$) $\Delta K_\gamma=G/4+\eta G^2$ (here we used $\eta \approx -0.000067$~m.  For $\eta =0 $ the result is similar, but with the pump wave remaining depleted for a shorter distance).
Panels ($c$) and ($f$) show the photon conversion efficiency (quantum yield of Eq.~\ref{quantumyield}) for the two optimum settings in both  CLPM configurations.} \label{DualCoreDistDep}
\end{figure}

The phase-modulation correction factors in Eqs.~(\ref{pmCorrDualalpha}-\ref{pmCorrDualgamma})  are valid when the pump wave at wavelength $\lambda_2$ is  injected in  a single core. We confirmed them numerically, from  the system of six coupled wave equations (Eqs.~(\ref{cw1}-\ref{cw3}) plus the corresponding set for the other waveguide), by checking that they do imply an energy transfer from the pump wave to the generated waves that reaches the theoretical maximum, and that is maintained over a wide range of pump  intensities. This was done for many  different configurations, with only  minimal adjustments of the small $\eta$ parameter. An example of the result of such numerical computations is shown in   Fig.\ref{DualCoreDistDep}.

The most important result here is that the $\Delta K_\alpha$ CLPM configuration   is insensitive to self- and cross-phase modulation. In other words, when this condition is fulfilled, the coupled-wave equations for the evolution of the wave amplitudes can always lead  to  the complete depletion of the pump wave, with no dependence from the intensity and no need for a phase-modulation correction term. This insensitivity to phase modulation is a unique advantage of this configuration, in addition to its potential 100\% quantum yield.

For the  $\Delta K_\gamma$ CLPM configuration, on the other hand, we found that phase modulation does have an effect, but that it can be taken into account in a similar way as for the single-mode fiber, this time using Eq.~\ref{corrPMKgamma}. We confirmed the validity of Eq.~\ref{corrPMKgamma}  up to the highest practical $G$-values of $100$~m$^{-1}$. Still, the  sensitivity to phase modulation is significantly attenuated also in this configuration when compared to the single core configuration, for which the phase modulation correction  is four times larger (this factor of four may be related to the fact that in this dual-core configuration the pump wave only occupies one of the cores for half the interaction length).

In practice, the plots in Fig.~\ref{CLPMwlDistance} are essentially unchanged  when substituting  Eq.~\ref{corrPMKgamma} for $\Delta K_\gamma = 0$. Even for $G=100$~m$^{-1}$, only the curves for a large distance between the cores are slightly modified (by a small displacement corresponding to a change in distance by 10\%). In fact, using Eq.~\ref{corrPMKgamma} for Fig.~\ref{CLPMrDwavelengths} does not lead to any important change in the curves plotted there, even for the relatively large value of $100$~m$^{-1}$ of the phase modulation correction term $G$. As an example, the core distance required to down-convert to $\lambda_1=2.2$~$\mu$m using the  $\Delta K_\gamma$ CLPM configuration and a core-radius of 4.5~$\mu$m changes by less than 1\% when taking into account phase modulation.

CLPM in dual-core fibers is therefore characterized by a strong,  intrinsic robustness against self- and cross-phase modulation, which have no effect at all on the conditions derived using the  $\Delta K_\alpha=0$ CLPM configuration, and only a small effect on those derived for the $\Delta K_\gamma=0$ CLPM configuration. In addition, we note that in a practical situation where the dual-core scheme is used to down-convert a single pump-wave, the effect of phase-matching is just to slightly shift the frequencies of the generated waves away from those calculated for a certain fiber geometry. For the example plotted in Fig.~\ref{CLPMwlDistance}, the choice of $\lambda_2 = 1.55$~$\mu$m and a distance between the cores of 10 $\mu$m leads to frequency downconversion to $\lambda_1=2.2$~$\mu$m, and phase-modulation only shifts this value by  less than 1\%.

The evolution of the interacting waves with propagation distance, as predicted by the coupled-wave equations (\ref{cw1}-\ref{cw3}), is depicted in  Fig.\ref{DualCoreDistDep}.

To better show the difference between these two  configurations for third-order wavelength down-conversion in a dual core fiber we show in Fig.\ref{DualCoreDistDep} the evolution of the normalized wave amplitudes $A_i$ in one of the two cores as a function of propagation length. We do that for a core radius of $3.5$~$\mu$m and a pump wavelength of $1.3$~$\mu$m for $\Delta K_\alpha=0$, and for a core radius of $4.5$~$\mu$m and a pump wavelength of $1.55$~$\mu$m for $\Delta K_\gamma=0$ and its modification of Eq.~(\ref{corrPMKgamma}). In both cases we kept the generated wavelength at $\lambda_1=2.2$~$\mu$m and adapted the distance between the cores to vary the values of $\Delta K\alpha$ and $\Delta K_\gamma$.

For this figure, we assumed an intensity of the pump wave similar to what we used in Fig.~\ref{singlemodesilicapropagation}, which gave a conversion length of a few meters, and therefore shorter than the absorption length for the generated wave near a wavelength of 2.2 $\mu$m, a region where the absorption of silica starts to grow rapidly.
The absorption of silica fibers at a $\lambda_1=2.2$~$\mu$m wavelength is of the order 100 dB/km \cite{Artyushenko14}, which gives a loss of 50\% after a propagation length of 30 m.  Even though long-wavelength absorption limits frequency downconversion in silica to shorter generated wavelengths,  the methodology developed above can be applied equally well to  any other kind of fiber. Examples are given in the next section.

\subsubsection{Fluoride dual core fibers}

The same kind of analysis that we used in the previous section can also be applied to fibers that transmit farther into the infrared such as those made from ZrF$_4$ or InF$_3$ glasses. 
For simplicity, and for better comparison with the case of silica fibers discussed above, we keep the same index contrast and core radius   of $4.5$~$\mu$m that we used for most of the silica examples. 

The main difference between these fluoride glasses and silica is well represented by the maximum pump wavelength that can be used to achieve phase matching in the single-core configuration. It corresponds to the zero group-velocity dispersion wavelength, and it increases from $\lambda_2^{max} = 1.248$~$\mu$m for silica, to $\lambda_2^{max} = 1.63$~$\mu$m for ZrF$_4$, to $\lambda_2^{max} = 1.91$~$\mu$m for the InF$_3$ example. This then has a clear effect on the wavelength-tuning CLPM conditions that we obtain for  ZrF$_4$ and InF$_3$, when compared to the corresponding curves plotted for silica in    Fig.~\ref{CLPMwlDistance}. 

\begin{figure}[h!]
  		 \centerline{\includegraphics[width=\linewidth]{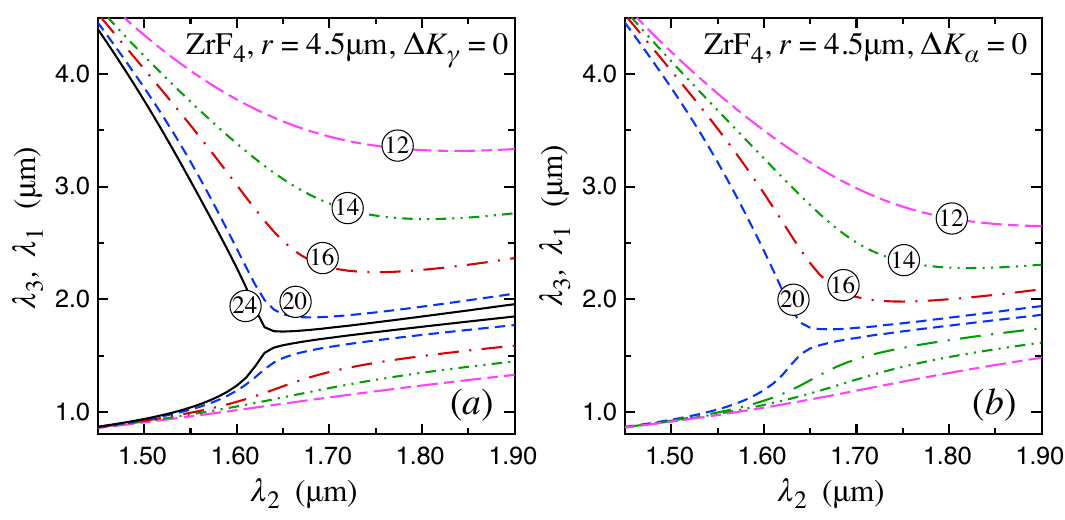} }
\caption{Wavelength-tuning CLPM curves  for different distances between the cores in a ZrF$_4$ dual core step-index fiber with a refractive index contrast of $\Delta n=0.01$ and core radius of 4.5~$\mu$m. Left: $\Delta K_\gamma=0$. Right: $\Delta K_\alpha=0$. The wavelength-tuning curves are labeled with the center-to-center distance between the cores in micrometers.} \label{CLPMwlDistanceZrF}
\end{figure}

The wavelength-tuning CLPM curves for $\Delta K_\alpha$ and $\Delta K_\gamma$ in a ZrF$_4$ fiber are shown in Fig~\ref{CLPMwlDistanceZrF}. The  core radius of $4.5$~$\mu$m increases the coupling for the generated wavelength $\lambda_1$ and creates larger deviations from the single-core limit as the two cores move closer to each other. For this configuration, the down-converted wave with wavelength $\lambda_1$ is always created in the even supermode, which will always exist. In this case, the small radius of $4.5$~$\mu$m makes it possible to still use a pump wavelength of $1.55$~$\mu$m to generate infrared light with wavelengths between $\lambda_1 \sim 3$~$\mu$m and $\lambda_2 \sim 4$~$\mu$m in the ZrF$_4$ fiber when varying the core distance between 20 and 12 $\mu$m.  

The situation in In$F_3$ fibers is qualitatively the same, but the maximum pump wavelength for single-mode interaction in this material is significantly larger, and it is not possible anymore to use either $\Delta K_\alpha$ or $\Delta K_\gamma$ to generate waves that are still in the transparency range of the material when using a pump wavelength of $1.55$~$\mu$m. In addition, even if they were not absorbed, the wavelengths of the generated waves would also be quite insensitive to changing the distance between the cores. 

\begin{figure}[h!]
  		 \centerline{\includegraphics[width=\linewidth]{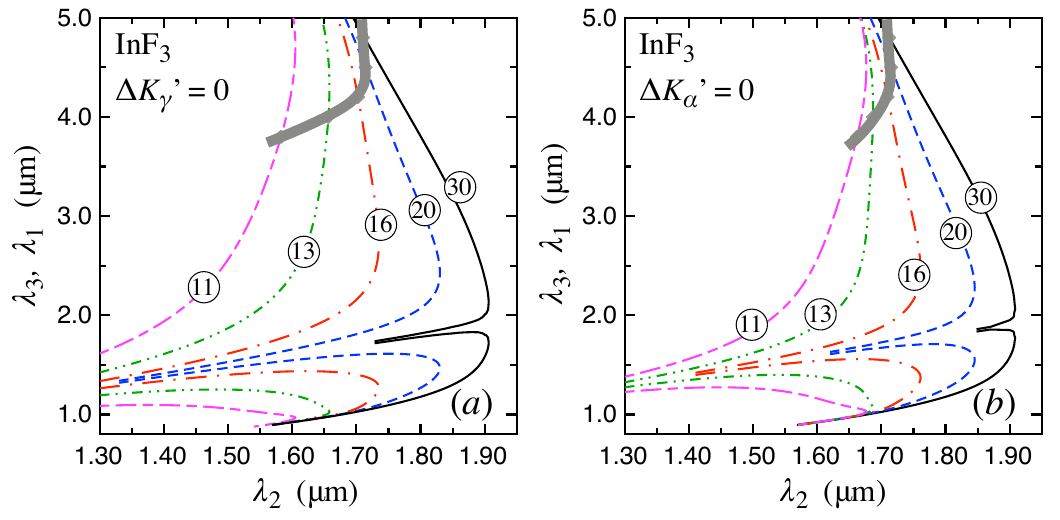} }
\caption{Wavelength-tuning CLPM curves  for different distances between the cores in a ZrF$_4$ dual core step-index fiber with a refractive index contrast of $\Delta n=0.01$ and core radius of 4.5~$\mu$m. Left: $\Delta K_\gamma '=0$. Right: $\Delta K_\alpha '=0$. The wavelength-tuning curves are labeled with the center-to-center distance between the cores in micrometers. For these two CLPM conditions the longest wavelength generated wave is created in the odd supermode; the thick gray lines towards the top of the graphs are the boundary above which the odd mode for the $\lambda_1$ disappears for the given distance between the cores.} \label{CLPMwlDistanceInF}
\end{figure}

However, the large  $\lambda_2^{max} = 1.91$~$\mu$m for In$F_3$ also means that pump wavelengths around 1.55~$\mu$m will be in a region where $\Delta k <0$ and then it becomes possible to obtain phase matching using $\Delta K_\alpha '$ and $\Delta K_\gamma '$. We plot these wavelength-tuning curves here, even though they are of limited use,  because they are qualitatively different from what we have seen up to now, with the generated wavelength $\lambda_1$ decreasing for decreasing distance between the cores. This is shown in Fig.~\ref{CLPMwlDistanceInF}. We note the small tunability interval between $\lambda_1 \approx 2$~$\mu$m   $\lambda_1 \approx 3$~$\mu$m  that can be achieved in this way in the  In$F_3$ fiber. This, coupled with the fact that the generated longest-wavelength wave is created in an odd supermode, makes  In$F_3$ an inferior choice when compared to  Zr$F_4$.

\section{ Conclusions}
We provided a full description of how a phase-matched interaction between propagating waves at three widely different frequencies can be achieved in dual core fibers, and we have shown that cross- and self-phase modulation can either be taken into account, or that its  effects  in dual-core fibers can  be fully eliminated for a specific CLPM configuration, and substantially attenuated otherwise. Because of this, the direction of energy transfer between the interacting waves can be theoretically maintained over indefinite fiber lengths, and it becomes possible to always approach the maximum conversion efficiency.

Nonlinear optical cross- and self-phase modulation effects, being always phase-matched and  independent from wavevector mismatch $\Delta k$,  can  give rise to the accumulation of an additional, intensity-dependent phase shift    between the interacting waves during propagation. When pumping a single-mode fiber with two waves and therefore defining the wavelengths involved in the interaction,   nonlinear phase-modulation will always disrupt any phase-matched energy transfer process between the waves. However, this can be counteracted by  adding an appropriate intensity-dependent term to the phase matching conditions, both for a single-core configuration and for a dual-core configuration.   For the case when there is only one pump wave and the generated waves arise from noise,  the wavelengths that offer the best quantum yield will be favored, and the nonlinear phase-modulation will result in a very small shift in the generated wavelengths away from the prediction dictated only by phase matching. A key result that we have obtained as part of our studies of the dual-core fiber system in this work, is that there is a specific CLPM condition that is completely insensitive to self- and cross-phase modulation.

While the focus of this work has been on frequency downconversion towards the farther infrared, all the results automatically apply  to the use of the same three-wave mixing scheme for upconversion and the generation of shorter wavelengths, or for difference-frequency generation or sum-frequency generation schemes using  third-order optical nonlinearities.

The optimum wavelength choices for the three-wave interaction that we discussed  can be tuned by varying index contrast, core radius, and the separation between the cores in dual core fibers. We have explored a few of the many possibilities. Most of the examples we provided  generally used a refractive index contrast of 0.01, a core radius of $4.5$ $\mu$m, and a pump wavelength of 1.55 $\mu$m, but variations of these  parameters will  further modify the phase-matching conditions, and design parameters and phase matching wavelengths are naturally also determined by the natural refractive index dispersion of the material used,  as described by the wavelength dependence of the single-core zero group-velocity dispersion. The rules and approach laid out in this work can be used in a straightforward way to design dual-core fibers with different index contrast and to further extend the examples that we provided in this work. A simple computer program can be used to vary the design parameters to numerically find the zeros in all CLPM conditions.

We provided a set of examples for using dual-core silica fibers pumped with 1.55 $\mu$m ($e.g$ from an Erbium fiber laser)  or 1.3 $\mu$m  wavelengths ($e.g.$ from a praseodymium fiber laser) to generate  infrared wavelengths beyond 2~$\mu$m, only limited towards longer wavelengths by silica absorption. This was followed by similar examples for dual-core fiber designs that can work as frequency converters in fluoride fibers, using the same pump wavelengths of 1.3 $\mu$m and 1.55 $\mu$m to generate farther infrared wavelengths around 4 $\mu$m.

When we provided an example of the dependence of the wave amplitudes on the propagation length, we chose a pump intensity about 2 orders of magnitude smaller that the damage threshold of silica \cite{Zervas14}, and we found that full conversion can be obtained after propagation distance of the order of a few meters or less. In general the conversion length is inversely proportional to the intensity of the pump wave. 

Finally, all our calculations were for an ideal, straight step-index fiber. Our examples can still serve as a guideline in general, but it is clear that new calculations will be needed for other situations or to take into account additional effects, as an example when the coupling constants between the cores (and the corresponding linear combinations of coupling constants that determine the CLPM conditions) are affected by the bending radius of longer fibers used for lower intensity pumps.

\begin{backmatter}

\bmsection{Disclosures} The authors declare no conflicts of interest.

\end{backmatter}

\bibliography{nlo}

\bibliographyfullrefs{nlo}


\ifthenelse{\equal{\journalref}{aop}}{%
\section*{Author Biographies}
\begingroup
\setlength\intextsep{0pt}
\begin{minipage}[t][6.3cm][t]{1.0\textwidth} 
  \begin{wrapfigure}{L}{0.25\textwidth}
    \includegraphics[width=0.25\textwidth]{john_smith.eps}
  \end{wrapfigure}
  \noindent
  {\bfseries John Smith} received his BSc (Mathematics) in 2000 from The University of Maryland. His research interests include lasers and optics.
\end{minipage}
\begin{minipage}{1.0\textwidth}
  \begin{wrapfigure}{L}{0.25\textwidth}
    \includegraphics[width=0.25\textwidth]{alice_smith.eps}
  \end{wrapfigure}
  \noindent
  {\bfseries Alice Smith} also received her BSc (Mathematics) in 2000 from The University of Maryland. Her research interests also include lasers and optics.
\end{minipage}
\endgroup
}{}

\end{document}